\documentclass[journal,twocolumn]{IEEEtran}
\usepackage{amsmath,amssymb,mathtools}
\usepackage{color}
\usepackage{hyperref}
\usepackage{graphicx}
\usepackage{csquotes}
\usepackage{cite}
\usepackage{epstopdf}
\usepackage{float}
\usepackage{url}
\usepackage{multirow}
\begin{document}
\title{\textit{eSampling}: Rethinking Sampling with Energy Harvesting}

\author{\IEEEauthorblockN{Neha Jain\textsuperscript{$\ast$,$\dagger$}, Bhawna Tiwari, Anubha Gupta\textsuperscript{$\dagger$}, Vivek Ashok Bohara\textsuperscript{$\ast$} and  Pydi Ganga Bahubalindruni\textsuperscript{$\ddagger$}}\\
\IEEEauthorblockA{\textsuperscript{$\ast$}Wirocomm Research Lab, \textsuperscript{$\dagger$}Signal processing and Bio-medical Imaging Lab (SBILab)\\
Department of Electronics and Communication Engineering, \\
Indraprastha Institute of Information Technology-Delhi (IIIT-D), New Delhi, India\\
\textsuperscript{$\ddagger$}Indian Institute of Technology, Goa, India\\
Email:nehaj@iiitd.ac.in, bhawnat@iiitd.ac.in, anubha@iiitd.ac.in, vivek.b@iiitd.ac.in and bpganga@iitgoa.ac.in }}



\maketitle
\begin{abstract}
In general, real world signals are analog in nature. To capture these signals for further processing, or transmission, signals are converted into digital bits using analog-to-digital converter (ADC). In this conversion, a good amount of signal energy is wasted because signal that is captured within the sampling duration is utilized, while rest of the signal waveform is discarded. In this context, this paper revisits the sampling process and proposes to utilize this discarded signal for energy harvesting, naming the method as \textit{eSampling}, i.e., sampling with energy harvesting. The proposed idea of \textit{eSampling} is demonstrated via modifying the circuitry of the hold phase of ADC. The system is designed using standard Complementary Metal Oxide Semiconductor (CMOS) 65 nm technology and simulations are performed on Cadence Virtuoso platform with input signal at different frequencies (100 Hz and 40 MHz). These results show that 10\% of the sampling period is sufficient to sample the input analog signal, while the remaining 90\% can be used for harvesting the energy from the input analog signal. In order to validate \textit{eSampling} for practical scenarios, results with  hardware setup have also been added.


\end{abstract}

\IEEEpeerreviewmaketitle

\section{Introduction}
Energy harvesting (EH) via ambient sources has become an important area of research due to the advent of low power \cite{7578025,1231858} and small form-factor devices \cite{371970,5402665}. Small form factor has not only led to reduction in size of electronic devices, but also enhanced it's functionality. This has also led to plethora of wearable devices that has replaced most of the functionalities of portable devices. For instance, smart watches have replaced some of the functionalities of smart phones. Further, the advancement in low power microelectronics technology has reduced the power consumption of devices by a considerable amount and hence, devices can be operated by harvesting the energy from ambient sources. Energy harvesting can either increase the lifetime of the device or can ensure a battery-free system \cite{8094977}. Similarly, in case of wireless sensor networks (WSNs), battery-free sensors can eliminate the difficulties of replacing or recharging the batteries at remotely located sensors \cite{7286751,6951347,basagni2013wireless}. EH also finds applications in toxic environments, medical body area networks (MBANs), and body area networks (BANs) \cite{4463339,6616846,6695200}, where replacing or recharging batteries is an arduous task. 

Energy can be harvested either from human activities such as walking, jogging, and cycling \cite{rao2013compact,starner2004human} or from environmental sources such as solar \cite{8402238, 4490281}, wind \cite{7748463}, thermal \cite{8444478} and mechanical pressure etc. \cite{5196774}. Recently, harvesting the energy from radio frequency (RF) signals has gained a lot of popularity over other energy sources \cite{6922609}. The foremost reason for the same is that RF signals, being harvested, are abundant in the humanized environment, whereas solar/wind resources are not perennially available. Furthermore, RF signals contain both information as well as energy and hence, an energy constrained receiver can decode the information as well as harvest the energy. This method is known as simultaneous information and power transfer (SWIPT). The two realistic protocols for SWIPT, namely, time switching (TS) and power splitting (PS) \cite{6489506,6503739} affect the system performance in order to harvest the energy. However, the phrase `energy harvesting' refers to the process of converting discarded/wasted/unused energy into useful electrical energy for fulfilling the energy requirements, for instance, solar EH. Thus, energy should be harvested from the discarded signals without degrading system performance.

In this paper, we propose a novel method for energy harvesting termed as  \textit{eSampling}, where \textit{e} signifies the process of energy harvesting. \textit{eSampling} has been proposed for harvesting the unused/discarded signal energy during the sampling process of an analog-to-digital converter (ADC). Since most of the real signals are analog in nature, to capture and store the data, these are converted into equivalent digital bits by using an ADC. An ADC consists of two processes, namely \textit{sampling} and \textit{quantization}. Sampling converts the input analog signal into discrete samples, while quantization converts the discrete samples into digital bits. Typically, a sample and hold (S/H) circuit is used to sample the input analog signal in an ADC. The S/H circuit consists of two phases, \textit{acquisition phase} and \textit{hold phase}. In acquisition phase, the S/H circuit tracks the input analog signal. The sampled value captured in the acquisition phase is converted into digital bits during hold phase. Therefore, during sampling process of an ADC, the input signal is processed only for a small fraction of time (acquisition phase) and neglected/discarded for the remaining time interval (hold phase). Hence, a method \textit{eSampling} is proposed which modifies the hold phase of the sampling process to harvest the energy from the input signal. To the best of our knowledge, no such work has been reported till date which harvests the input signal energy during the sampling process without degrading the performance of an ADC or any another block in the system. 

\subsection{Related work}
In literature, methods have been proposed for harvesting energy from discarded signals \cite{793,7145810}. For example, a method has been proposed for harvesting energy during modulation and filtration in \cite{793}. Since a part of signal energy is often discarded or lost during modulation, this can be harvested to generate electrical power. Likewise, filtering also discards some part of the signal in the process of accomplishing desirable task. Signal portion discarded during filtering can be utilized to harvest energy. Similarly, signal energy has been harvested from the redundant portion (i.e. the cyclic prefix) in Orthogonal Frequency Division Multiple Access (OFDMA) system in \cite{7145810}.

The method \textit{eSampling} has worldwide applications, since real signals are analog in nature and most of the existing systems are energy constrained. For example, the proposed method can be directly utilized in WSNs because the sensor nodes are energy constrained nodes and are typically powered by batteries with a limited lifetime \cite{8515268}. Therefore, energy harvesting systems can be employed to supplement the lifetime of the battery. Most of the recent work focuses on developing an effective energy management solution \cite{8620539, 7105895, 8586917,8362666} because even though additional energy can be harvested from the external environment, it remains a limited resource, and hence should be consumed wisely. Effective energy management relies on re-designing of circuit to harvest maximum available energy \cite{8620539,7105895} and also on re-designing of network which includes various opportunistic transmission strategies and routing algorithms to reduce the energy consumption \cite{8586917,8362666}. On the other hand, the current work proposes an entirely different approach and provides an alternative form of energy harvesting that is carried out during the sampling of its own sensed analog signal.  Generally, sensor nodes generate analog signals corresponding to the sensed parameters that are digitized using ADC and are transmitted to the Fusion center \cite{8515268,basagni2013wireless}. The generated analog signals at sensor nodes have a good amount of energy that was wasted earlier. This energy can be harvested by using the proposed \textit{eSampling} method. Therefore, self-sustainable wireless sensor nodes can be fabricated that can charge themselves from their own sensed parameter without using any dedicated energy harvesting source. Similar to above, \textit{eSampling} method will find tremendous utility in wireless body area network (WBAN) because sensor devices are generally autonomous and implanted inside the body, where batteries are irreplaceable \cite{7914600}. Further, the proposed \textit{eSampling} method can be applied to  numerous industrial application scenarios including  but  not  limited  to  radar,  wireless communication \cite{6105586,6341363,7598271}, vehicular networks \cite{7462487} and  Internet of things \cite{8620539, 7105895, 8586917,8362666}.  In  future, energy  constrained  devices  or  circuits  involving  sampling process  at  some stage  may  like  to  replace  the  conventional sampling with \textit{eSampling} with re-design of the associated circuity.

Rest of the paper is organized as follows: Section II describes the proposed \textit{eSampling} method, Section III reports the simulation model designed for \textit{eSampling}, Section IV discusses simulation results for the validation of the proposed method and Section V provides the real time validation of \textit{eSampling} using hardware setup. Finally, conclusions are drawn in section VI.

\section{Proposed \textit{eSampling} method}
\begin{figure*}[ht]
\centering
\includegraphics[scale=0.25]{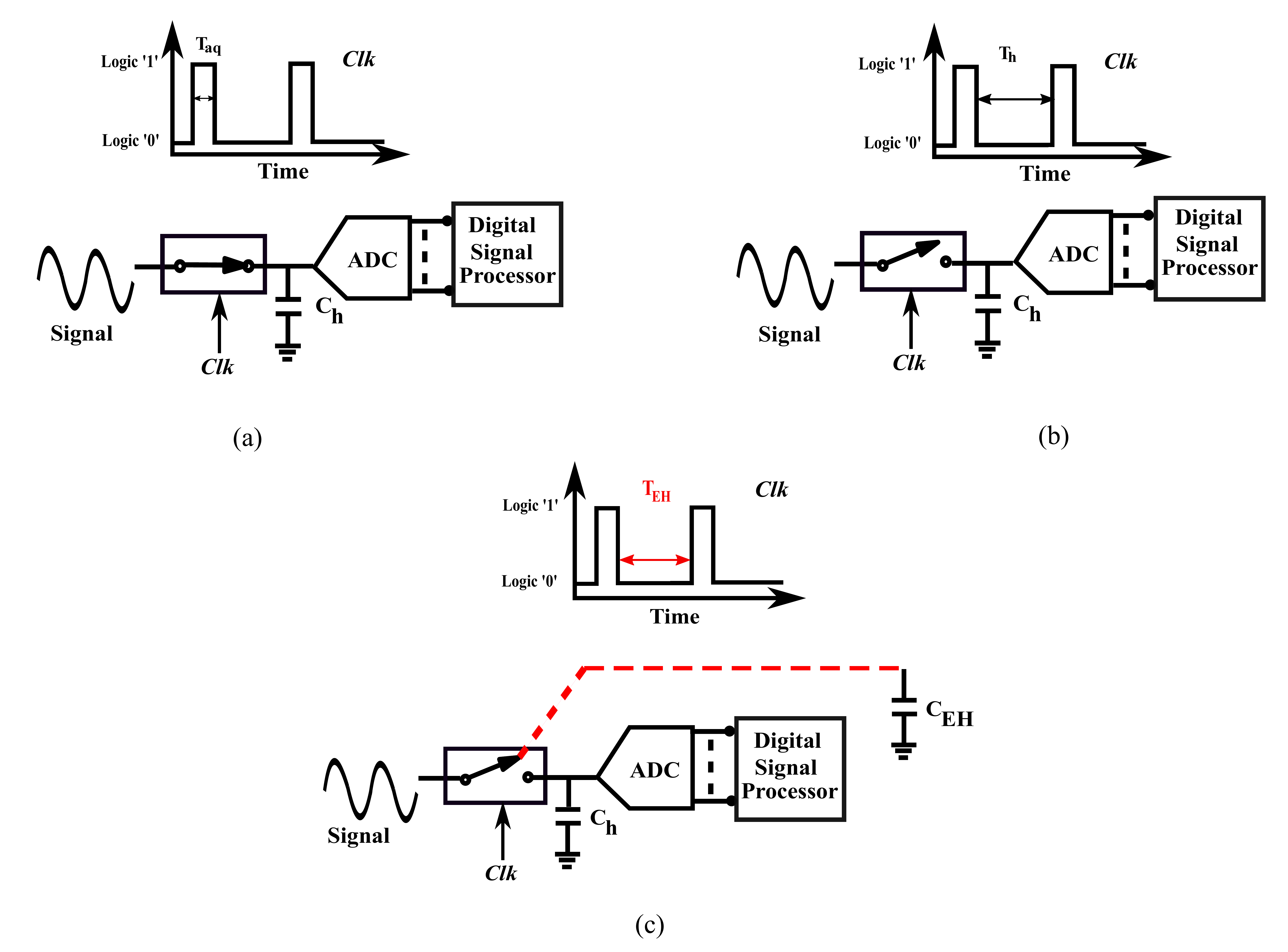}\caption{Typical data acquisition system with different phases (a) Acquisition phase (b) Hold phase and (c) Energy harvesting phase.}
\label{fig:1}
\end{figure*}

\begin{figure*}[ht]
\includegraphics[scale=0.2]{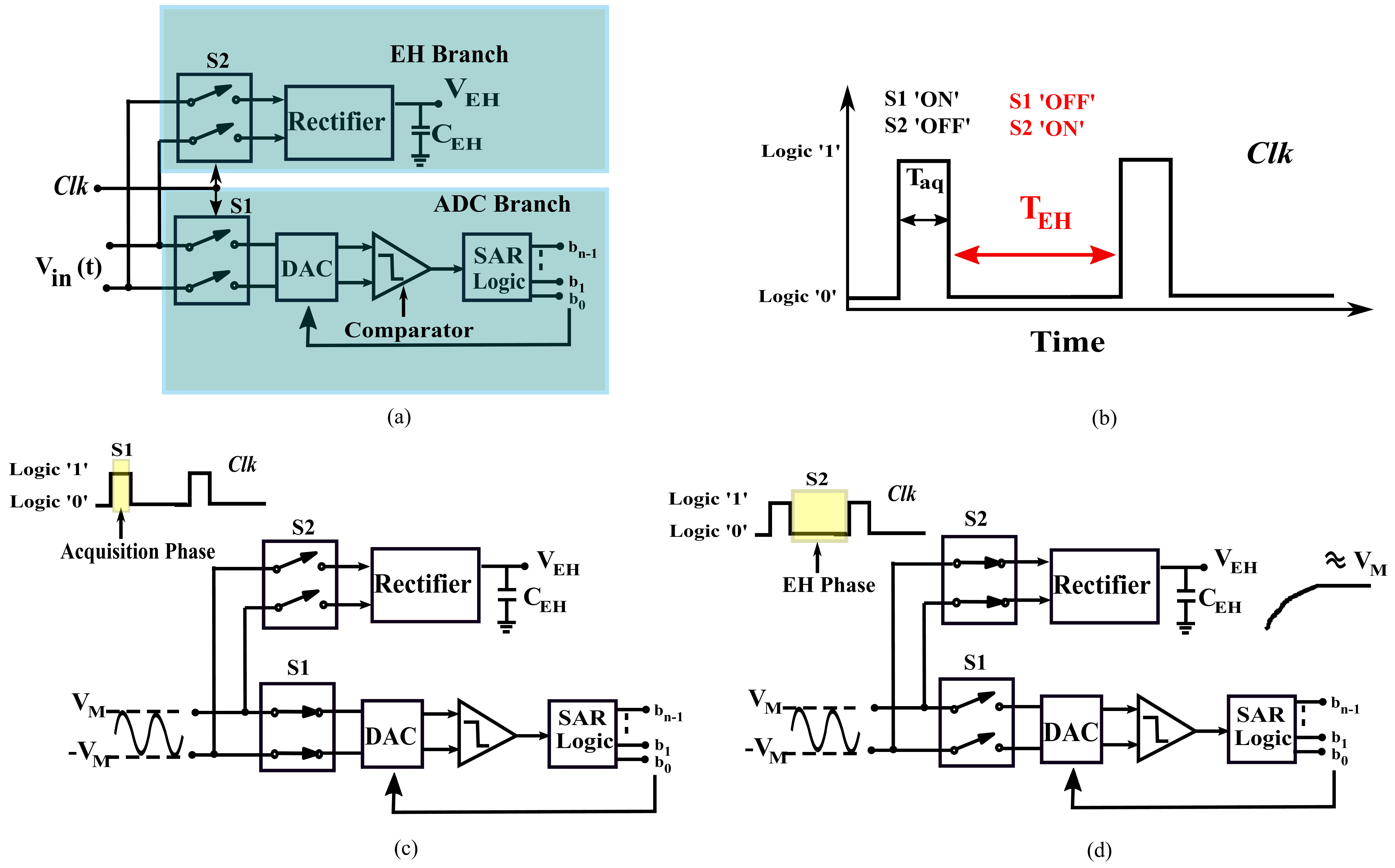}
\caption{Block diagram of (a) \textit{eSampling} system model, (b) Sampling clock, (c) Acquisition phase and (d) EH phase.} 
\label{fig:21}
\end{figure*}

\subsection{Overview}
Consider Fig. \ref{fig:1}a) and b) that describe the acquisition and hold phase of the sampling process, respectively controlled by a common sampling clock ($Clk$). The acquisition phase is the duration for which S/H switch is connected to capacitor $\mathrm{C_h}$ to track the input analog signal. Once the acquisition phase is over, the hold phase starts. In this phase, the S/H switch is open or disconnected from $\mathrm{C_h}$ as shown in Fig. \ref{fig:1}b) such that $\mathrm{C_h}$ can hold the sampled value acquired in the acquisition phase. This signal value is converted into digital bits using the quantization process. This is to be noted that the discarded input signal during the hold phase contains a considerable amount of energy that can be harvested. In this work, we propose an alternative phase, henceforth called as EH phase, during which the switch will be connected to another capacitor $\mathrm{C_{EH}}$ to store the energy as shown in Fig. \ref{fig:1}c). This phase can be seen as the modified version of the hold phase of the conventional ADC. It should be noted that the modified hold phase with the EH phase does not impact the functionality of the conventional ADC circuit.

Consider an input analog signal with maximum signal frequency $f_m$ sampled at Nyquist sampling rate. Hence, the sampling period is given as $T_s=\frac{1}{f_{s}}\leq\frac{1}{2f_m}$, where $f_s$ is the sampling frequency. The sampling period is given by
\begin{equation}
T_s = T_{aq}+T_h, \label{4}
\end{equation}
where $T_{aq}$ is the acquisition time or the time allocated to the acquisition phase as shown in Fig. \ref{fig:1}a) . It is defined as the minimum time required for $\mathrm{C_{h}}$ to charge itself to the full input signal voltage. $T_h$ is the hold time during which the input analog signal is disconnected from the ADC and the sampled value obtained across $\mathrm{C_{h}}$ is converted by ADC into corresponding digital bits as shown in Fig. \ref{fig:1}b). Since the input analog signal is not processed during $T_h$, it can be harvested to recharge the batteries or to drive other on-chip circuits.

\subsection{\textit{eSampling} System Model}
Fig. \ref{fig:21}(a) presents the block diagram of \textit{eSampling} system with input signal $(V_{in}(t))$. It comprises of two branches: ADC branch and EH branch. A differential Successive-Approximation Register (SAR) ADC has been employed in the ADC branch of the system that comprises of a sampling switch ($S1$), a voltage comparator, a digital-to-analog converter (DAC), and a SAR logic. On the other hand, EH branch consists of a switch $S2$, a rectifier, and a energy storage capacitor ($C_{EH}$). The voltage across the capacitor is denoted as $V_{EH}$. Switches $S1$ and $S2$ of the model are controlled by \textit{Clk} as shown in Fig. \ref{fig:21}(b).
The operation of \textit{eSampling} system is divided in two phases:\\

\textbf{Acquisition Phase} (S1 is ON and S2 is OFF): The time period of this phase is equal to $T_{aq}$. The equivalent block diagram is shown in Fig. \ref{fig:21}(c). The functioning of both ADC and EH branch in acquisition phase is described below.
\begin{figure*}[ht]
\centering
\includegraphics[scale=0.45]{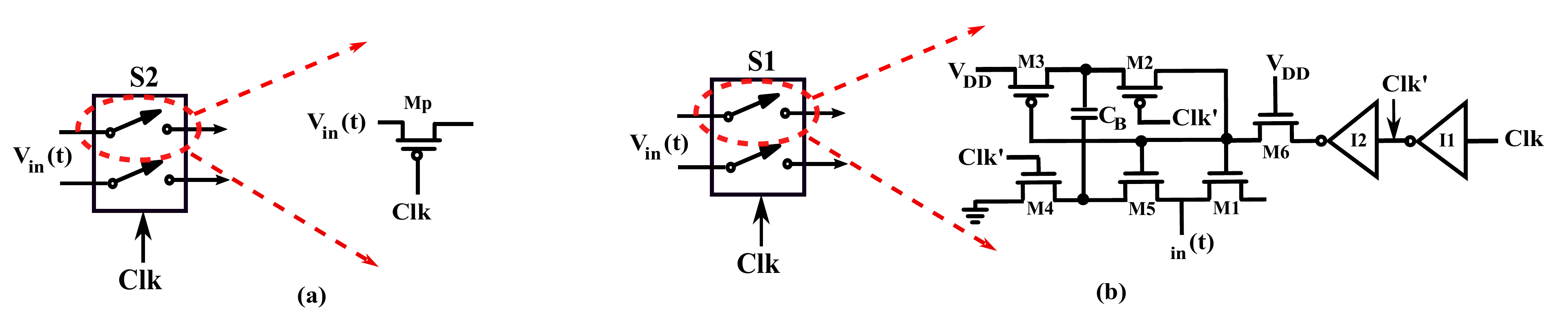}
\caption{Circuit Diagram of (a) p-channel pass transistor switch used as S2 in system model, (b) Bootstrapped switch used as S1 in system model.} 
\label{fig:23}
\end{figure*}
\begin{itemize}
\item ADC branch: Input analog signal from the surrounding or preceding blocks is sampled onto the DAC capacitors of SAR ADC \cite{hariprasath2010merged}.
\item EH branch: It is isolated from the signal source $(V_{in}(t))$.
\end{itemize}


\textbf{Energy harvesting (EH) Phase} (S1 is OFF and S2 is ON): The time period of this phase is equal to $T_{h}$. The equivalent block diagram is shown in Fig. \ref{fig:21}(d). The functioning of ADC and EH branch in EH phase is explained below.
\begin{itemize}
\item ADC branch: SAR ADC converts the sampled signal into equivalent digital bits using a quantizer.

\item EH branch: The input analog signal is rectified to store the maximum amplitude of the signal on capacitor $C_{EH}$. However, due to non-idealities associated with the rectifier and switches, certain voltage drop is experienced. Hence, the maximum voltage on $C_{EH}$ is given by 
\begin{equation}
    V_{EH}=\eta_v V_M, \label{1}
\end{equation}
where $V_M$ is the maximum amplitude of the input signal $V_{in}(t)$ and $0 < \eta_v \leq 1$ is the efficiency of the EH branch. Further, the maximum time that can be assigned for harvesting the input signal energy is given by using \eqref{4} as below:
\begin{align}
    T_{EH}&=T_{s}-T_{aq}. \label{2}
\end{align}

The amount of energy harvested/stored from the input analog signal can be calculated as
\begin{equation}
    E_h=\frac{1}{2}C_{EH}V^2_{EH}, \label{3}
\end{equation}
 The conversion efficiency, i.e., the percentage of energy harvested from the input signal until $C_{EH}$ attains the steady state is given by
\begin{align}
    \eta_e&=\frac{{E_h}}{E_{in}} \nonumber \\
    &=\frac{C_{EH}V^2_{EH}}{2P_{in}T_{C_{EH}}}, \label{5}
\end{align}
where $P_{in}$ is the root mean square (rms) power of the input signal $V_{in}(t)$ and ${T_{C_{EH}}}$ denotes the time taken by energy harvesting capacitor ($C_{EH}$) to charge up to steady state voltage, i.e., $V_{EH}$.

\end{itemize}




\section{Simulation Model}

In order to demonstrate the proposed idea, a system model is designed using standard  65 nm Complementary Metal Oxide Semiconductor (CMOS) technology. Design details of all sub-blocks in the system model are described below.

\subsubsection{Switches}
Switches $S1$ and $S2$ follow differential implementation and are realized using two different switch topology. $S1$ is implemented using n-channel MOS (NMOS) bootstrapped switch \cite{7258484}, while $S2$ is implemented using p-channel MOS (PMOS) pass transistors. Both switches are controlled by $Clk$.


Switch $S2$ is implemented using two PMOS transistors as shown in Fig. 3(a). The PMOS transistor turns ON and passes the input signal ($V_{in}(t)$) when $Clk$ is at logic '0'. However, it turns OFF and isolates the input signal from the next block when $Clk$ is at logic '1'. PMOS transistors are used instead of NMOS because PMOS transistors pass the logic `1' (high voltage) without experiencing the voltage drop. 

The on-resistance, $R_{ON}$ of a pass transistor $Mp$ is given by \cite{razavi1995principles}
\begin{equation}
R_{ON}=\frac{1}{\mu C_{ox}\frac{W}{L}(|V_{GS}|-|V_{TH}|)}, \label{12}
\end{equation}
where $\mu$ is the mobility,  $C_{ox}$ is the oxide capacitance, $\frac{W}{L}$ is the aspect ratio, $V_{GS}$ is the gate-to-source voltage, and $V_{TH}$ is the threshold voltage of the transistor. $V_{GS}$ of the pass transistor varies with time according to the input signal $V_{in}(t)$ and hence, $R_{ON}$ will also vary as given in  \eqref{12}. The variation in the $R_{ON}$ introduces non-linear distortion at the output of ADC and hence, bootstrapped switch is generally used for ADC. The bootstrap switch can typically ensure constant $R_{ON}$ \cite{7258484}.

Switch $S1$ is implemented using two NMOS single-ended bootstrapped switches as shown in Fig. 3(b). In order to ensure a nearly constant $R_{ON}$, the gate of transistor $M1$ is bootstrapped using two PMOS transistors $M2$ and $M3$, three NMOS transistors $M4,~M5$ and $M6$, and one capacitor $C_{B}$ \cite{7258484}. Two CMOS inverters $I1$ and $I2$ are also employed in the structure to generate required clock signals for proper operation of the switch.

\subsubsection{ADC}
The architecture of SAR ADC is implemented using a sampling switch $S1$, voltage comparator, SAR logic, and DAC. Switch $S1$ is a bootstrap switch as explained above. A two stage voltage comparator is implemented, where the first stage is a dynamic pre-amplifier and second stage is a latch \cite{8345180}. Furthermore, SAR logic is implemented using two arrays of shift registers that operate in serial-in-serial-out and parallel-in-parallel-out modes \cite{5771068}. The ADC operates asynchronously, hence the clock pulses for comparator and SAR logic are generated internally from $Clk$ using clock generation logic \cite{5771068}. In addition, binary-weighted capacitive DAC is used in the ADC. The capacitive DAC is implemented using merge capacitor switching (MCS) scheme \cite{hariprasath2010merged} because it eliminates the most significant bit (MSB) capacitor compared to conventional design \cite{1050630} and hence, reduces the total capacitance of the DAC ($C_{DAC}$) by half. It should be noted that $C_{DAC}$ of SAR ADC mimics $C_h$ (Fig. \ref{fig:1}). By reducing $C_{DAC}$, $T_{aq}$ of the ADC can be reduced, because $T_{aq}$ is defined as \cite{razavi1995principles}
\begin{align}
   T_{aq}& = R_{ON}^{S1} C_{DAC}, \label{20}
\end{align}
where $R_{ON}^{S1}$ is the on-resistance of $S1$ switch. Further, $C_{DAC}$ for $n$-bit SAR ADC is defined as
\begin{equation}
    C_{DAC}=\left ( 1+ \sum_{i=0}^{n-2}2^i \right )C_u,
\end{equation}
where $C_u$ is the unit capacitor of the DAC array. \textit{eSampling} demands low value of $T_{aq}$ so that more time could be allocated for harvesting the input signal energy (\ref{2}). It should be noted that to further reduce $T_{aq}$, $C_{DAC}$ and $R_{ON}^{S1}$ should be reduced as evident from (\ref{20}). Reducing $R_{ON}^{S1}$ demands wider transistors, which in turn increases the device capacitance and thus, reduces the operating speed of the ADC. In addition, wider devices leads to charge injection problem, which degrades the SNR of the ADC and hence, the performance of the ADC is compromised. On the other hand, employing small values for $C_{DAC}$ results in mismatch issues and sampling noise, which degrade the ADC conversion accuracy. These issues demand the $T_{aq}$ to be large enough such that the ADC performance is not compromised. A large $T_{aq}$ will reduce time available for the energy harvesting, $T_{EH}$. As a consequence, there is trade-off between $T_{EH}$ and $T_{aq}$.

\subsubsection{Rectifier}
\begin{figure}[ht]
\includegraphics[scale=0.32]{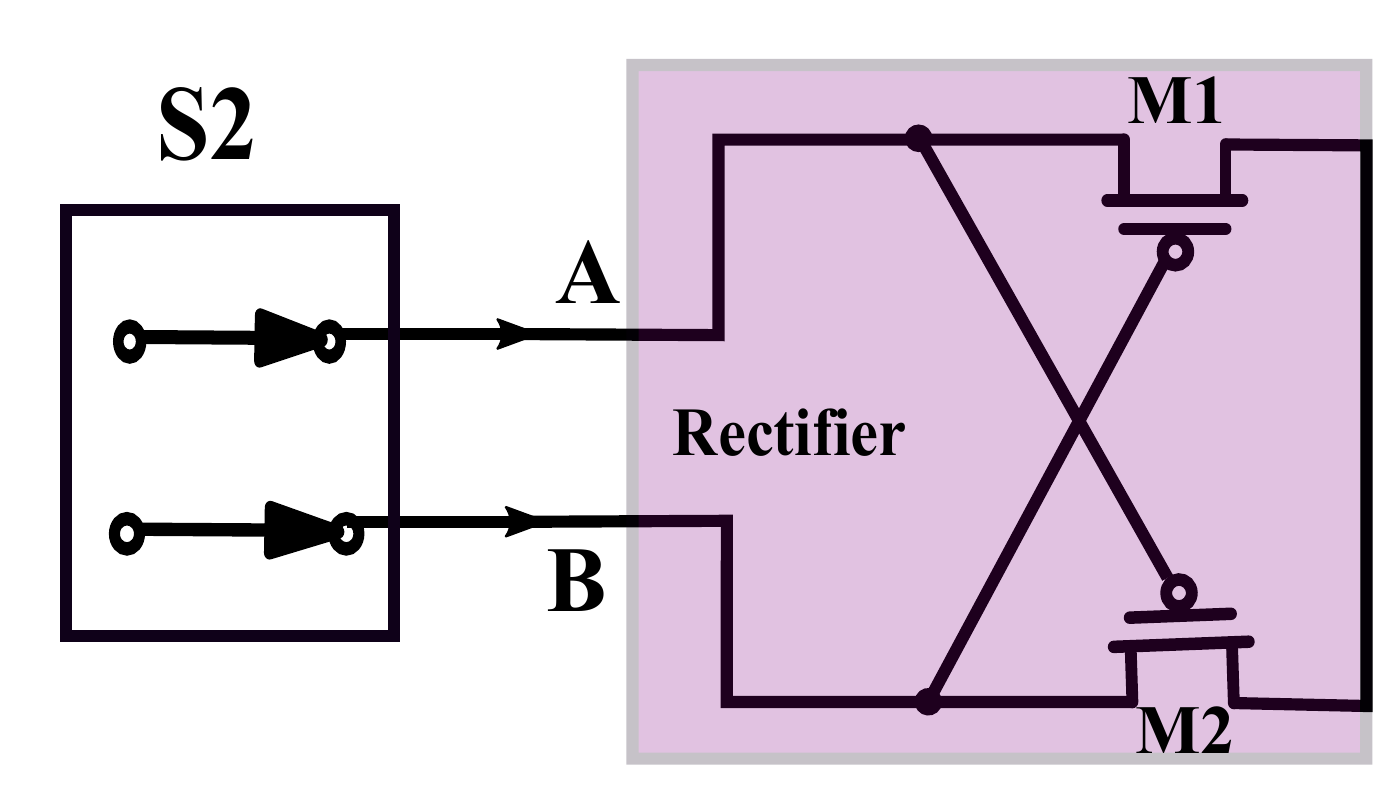}
\centering
\caption{Circuit Diagram of Rectifier.} 
\label{fig:241}
\end{figure}

A full wave rectifier has been employed in the simulation model as shown in Fig. \ref{fig:241}. Due to inherent threshold voltage drop in NMOS transistor, the rectifier is implemented using PMOS (p-type MOS) transistors. The two PMOS transistors have been connected in a cross-coupled fashion as shown in Fig. \ref{fig:241}. It should be noted that in this circuit, a negative to positive transition at point \textbf{A} results in positive to negative transition at point \textbf{B}, and hence, $M1$ is ON and $M2$ is OFF. While in vice-versa condition, $M1$ is OFF and $M2$ is ON. Therefore, a full wave rectification can be observed at the output.

\section{Simulation Results and Discussions}
\begin{figure}[ht]
\includegraphics[scale=0.32]{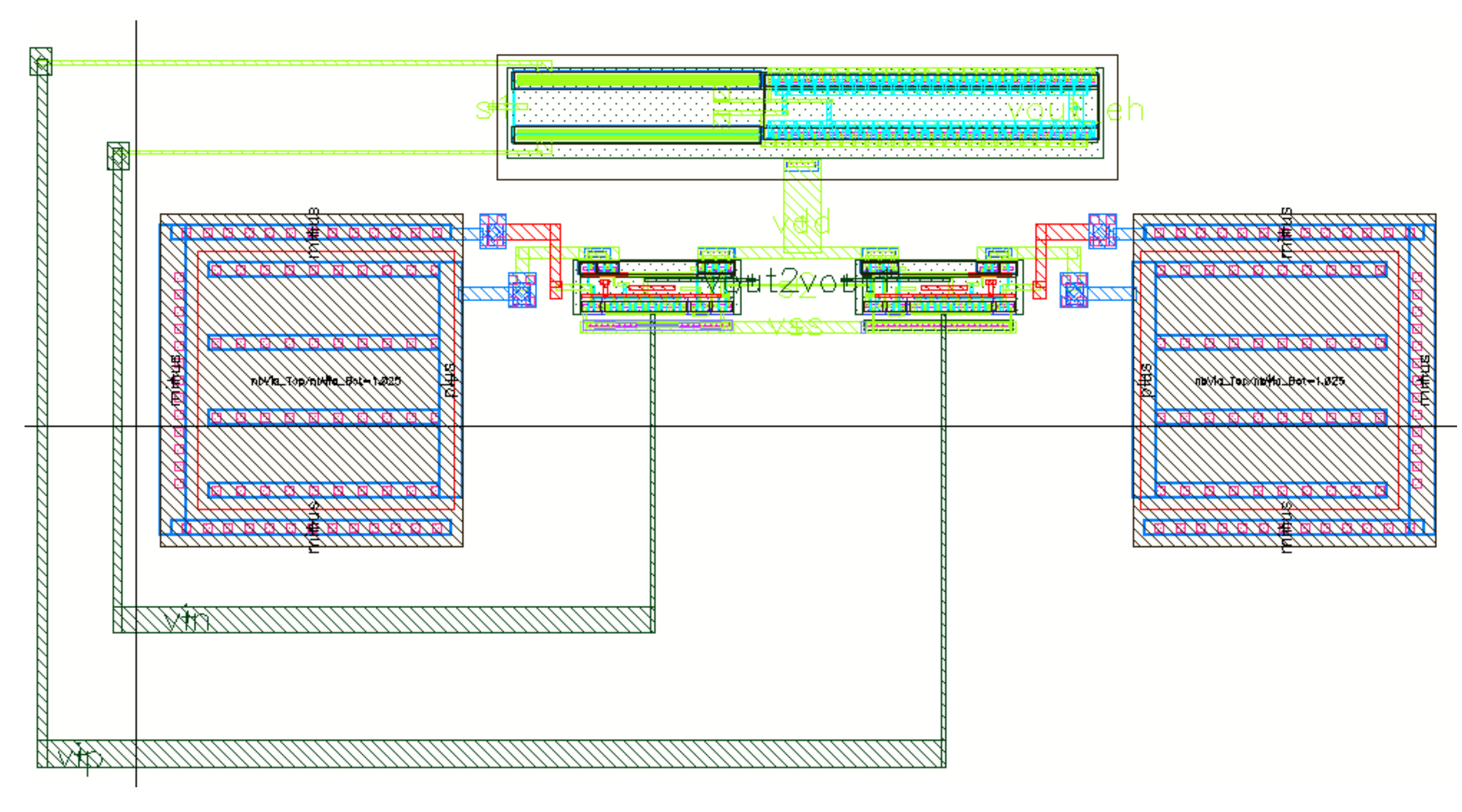}
\caption{Layout of the \textit{eSampling} model.}  
\label{fig:lay}
\end{figure}

A complete schematic of the \textit{eSampling} model has been created in Cadence Virtuoso platform. Schematic simulation uses mathematical models, which help us to authenticate the behavior of \textit{eSampling}. However, schematic simulations exclude the effect of parasitic capacitance and resistance associated with the connecting wires and devices employed in the system model. Therefore, post-layout simulations have been also performed to ensure similar performance of the model when fabricated to work in real world environment. Since the focus of the paper is to harvest maximum energy from the input signal without effecting the performance of ADC, post-layout simulations were carried out for switch \textit{S1} and EH branch. Layouts of these blocks are presented in Fig. \ref{fig:lay}. It should be noted that for obtaining the desirable performance of $n$-bit ADC, the output signal-to-noise and distortion ratio (SNDR) of \textit{S1} should be greater than $6.02n+1.76$ dB \cite{razavi1995principles}.

To demonstrate the efficiency of the proposed idea of energy harvesting along with sampling, an 8-bit SAR ADC has been used in the ADC branch of the \textit{eSampling} system. In addition, to cover a wide range of applications, \textit{eSampling} system has been simulated for both low (KHz) as well as high (MHz) input signal frequencies. During simulation, the acquisition time is chosen to be 10\% of the total sampling period, i.e., $T_{aq}= 0.1~ T_{s}$ such that 90\% of the sampling time can be allocated for energy harvesting, i.e., $T_{EH} \approx 0.9 ~T_{s}$. 

\subsection{Low-frequency Simulation of \textit{eSampling} System}
Low frequency simulations are performed at a sampling frequency $(\frac{1}{T_{s}})$ of 10 KHz. The frequency of the input signal is calculated using coherent sampling principle and following the Nyquist rule. A sinusoidal signal with amplitude varying from -0.4 V to 0.4 V is selected as an input with an input rms power of 27.7 $\mu$W for a source impedance of 50 ohm.

\begin{figure}[ht]
\centering
\includegraphics[scale=0.4]{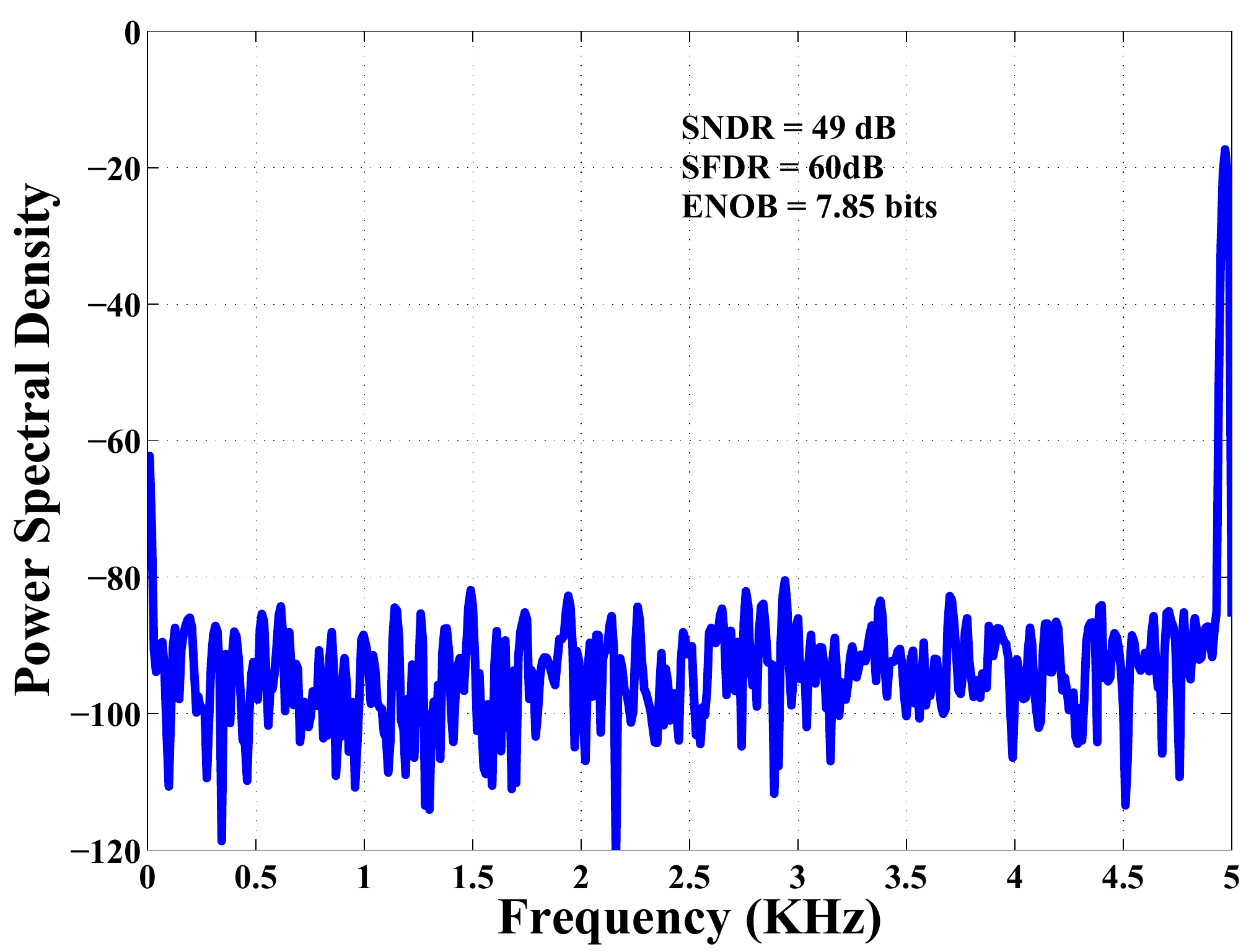}
\caption{FFT plot of 8-bit SAR ADC for 10 KHz sampling frequency.} 
\label{fig:29}
\end{figure}

A SAR ADC has been designed to operate at $T_{aq} = 0.1~T_s = 10 ~\mu$s with $C_u~=~12$ nF. From schematic simulations, the output SNDR of $S1$ is around 65 dB. The FFT plot of the reconstructed signal is presented in Fig.~\ref{fig:29}, which shows SNDR of 49 dB and ENOB (equivalent number of bits) equals to 7.85 bits for an 8 bit ADC that implies no loss of bits. In addition, from post-layout simulations, the obtained output SNDR of $S1$ is around 62 dB, which is greater than $6.02n+1.76$ dB. Therefore, as mentioned above, it can drive an 8-bit SAR ADC. 

\begin{figure}[ht]
\includegraphics[scale=0.4]{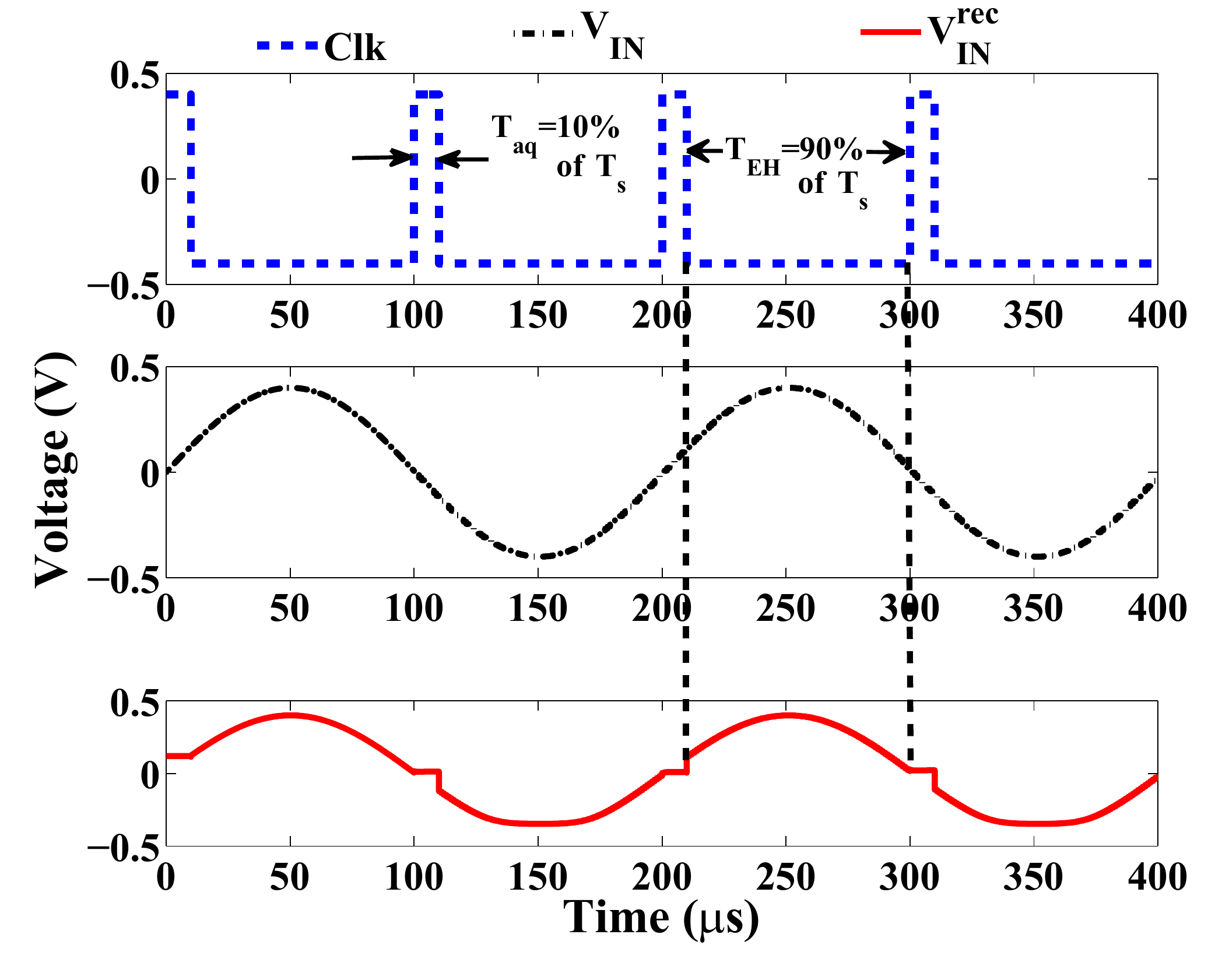}
\caption{Sampling clock, input signal and input to the rectifier in the EH branch at 10 KHz sampling frequency.}  
\label{fig:261}
\end{figure}

\begin{figure}[ht]
\includegraphics[scale=0.4]{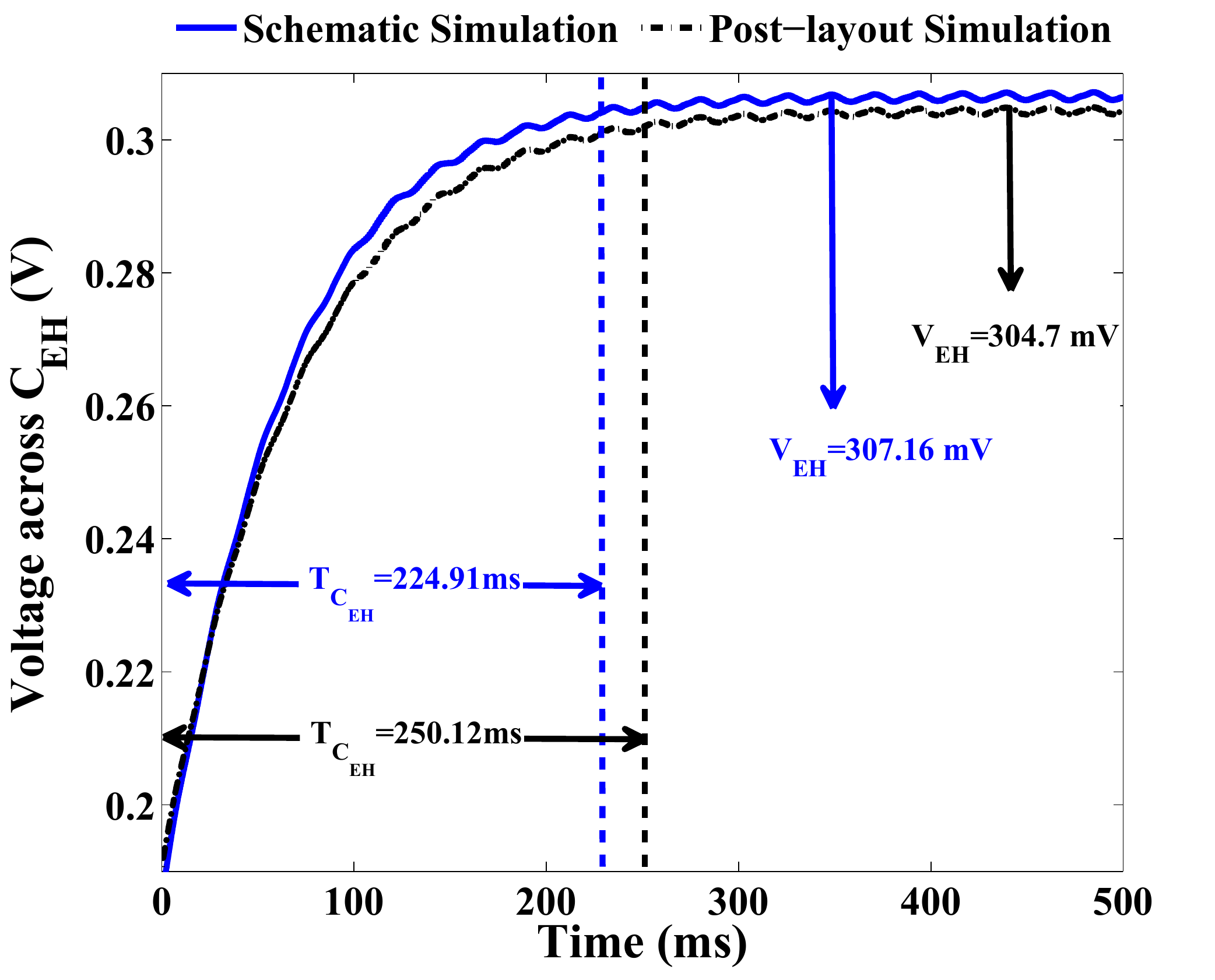}
\caption{Voltage obtained across $C_{EH}$ at 10 KHz sampling frequency.}  
\label{fig:26}
\end{figure}

Simulation results of EH branch is presented in Fig.~\ref{fig:261}. It is observed that when \textit{Clk} is `logic 0' (during $T_{EH}$ time period), the two PMOS transistors defining the $S2$ switch track the input analog signal $V_{in}(t)$ that will be the input of the rectifier $V_{in}^{rec}$. Rectifier rectifies $V_{in}^{rec}$ to provide a DC output voltage $V_{EH}$. Consider Fig. \ref{fig:26} for both schematic and post-layout simulation results for $V_{EH}$.

For schematic simulation a DC voltage close to 307.16 mV is obtained at $C_{EH} =100~\mu$F. This implies $\eta_v=76.79\%$. The time taken by the capacitor to charge up to 307.16 mV is $T_{C_{EH}} = 224.91$ ms. Therefore, $75.7\%$ of the input signal energy has been harvested by using the \textit{eSampling} system. In addition, $V_{EH}$, $T_{C_{EH}}$, $\eta_v$, and $\eta_e$ have been calculated from post-layout simulations as 304.7 mV, 250.12~ms, 76.175\% and around 67\%, respectively.  The degradation in performance is due to the parasitic capacitance and resistance involved in the post-layout simulations.

\subsection{High-frequency Simulation of \textit{eSampling} System}
In this section, the proposed model is simulated at a sampling frequency of 40 MHz. Following the above specification, the SAR ADC is designed to operate at $T_{aq} = 0.1~ T_s = 2.5$ ns with $C_u~=~15$ fF.  A sinusoidal signal with $V_M=0.4$ V is considered with the source impedance of 50 Ohm. The obtained input signal rms power $(P_{in})$ is 27.255 $\mu$W.

\begin{figure}[ht]
\centering
\includegraphics[scale=0.4]{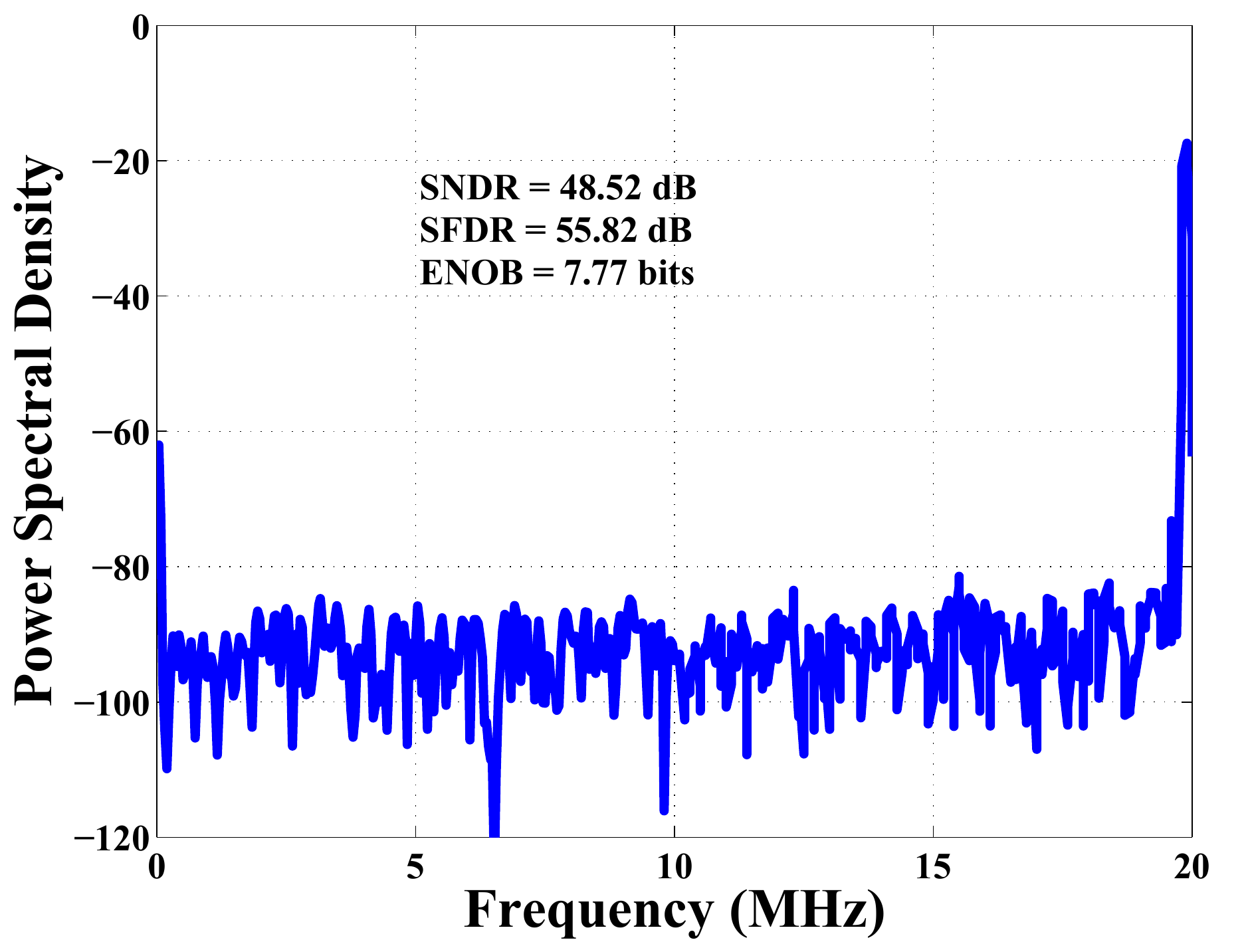}
\caption{FFT plot of 8-bit SAR ADC for 40 MHz sampling frequency.} 
\label{fig:28}
\end{figure}

From schematic simulations, the output SNDR of $S1$ is around 61 dB. The FFT plot of the reconstructed signal is presented in Fig.~\ref{fig:28}. From the plot, it is observed that the achievable SNDR and ENOB are 48.52 dB and 7.77 bits, respectively. Furthermore, from post-layout simulations, the obtained output SNDR of $S1$ is around 58 dB, which is greater than $6.02n+1.76$ dB. Therefore, as mentioned above, it can drive an 8-bit SAR ADC. 

\begin{figure}[ht]
\includegraphics[scale=0.4]{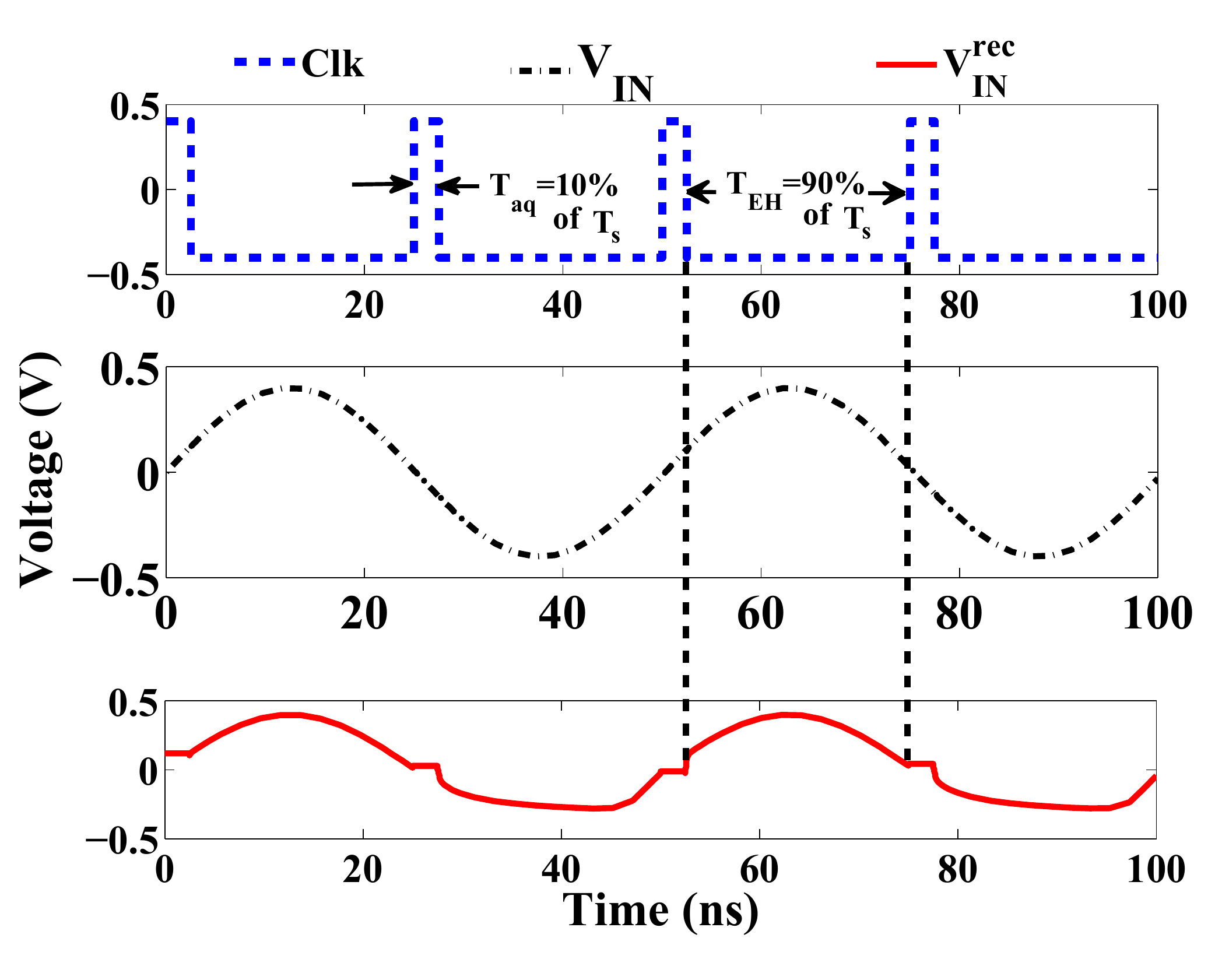}
\caption{Sampling clock, input signal and input to the rectifier in the EH branch at 40 MHz.} 
\label{fig:271}
\end{figure}

\begin{figure}[ht]
\includegraphics[scale=0.4]{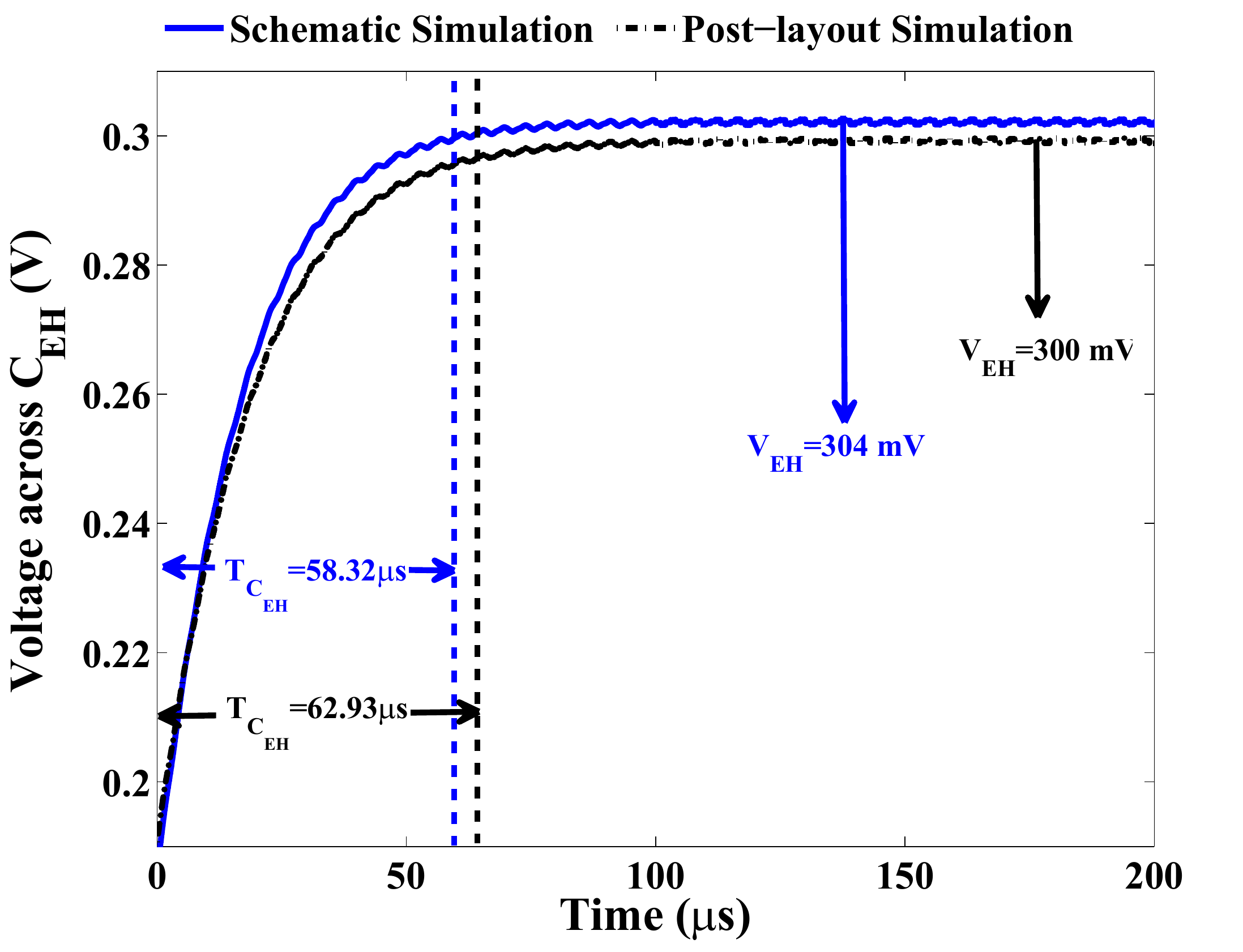}
\caption{Voltage obtained across $C_{EH}$ at 45 MHz KHz sampling frequency.}
\label{fig:27}
\end{figure}

Simulation results of EH branch is presented in Fig.~\ref{fig:271}. Compared to the low frequency operation (Fig. \ref{fig:261}), less time is available for energy harvesting during the high frequency operation. The obtained $V_{in}^{rec}$ is rectified and stored in $C_{EH}$ as shown in Fig. \ref{fig:27}. From schematic simulations, a DC voltage of 304 mV is obtained across $C_{EH} = 25$ nF without degrading the performance of ADC. This implies $\eta_v=76\%$. The time taken by $C_{EH}$ to charge up to $V_{EH}=304$ mV is $T_{C_{EH}}=58.32~\mu$s. Therefore, $72.64\%$ of the input energy could be harvested using the proposed system model at high frequency. In addition, $V_{EH}$, $T_{C_{EH}}$, $\eta_v$, and $\eta_e$ have been calculated from post-layout simulations as 300 mV, $63~\mu$s, 75\% and around 65.53\%, respectively. The degradation in performance is due to the parasitic capacitance and resistance involved in the post-layout simulations.

\begin{table*}[]
\centering
\caption{Performance Metric}
\label{tab:my-table}
\begin{tabular}{|c|c|c|c|c|c|c|c|c|c|}
\hline
\multicolumn{2}{|c|}{\textbf{Operation}} & \begin{tabular}[c]{@{}c@{}}$T_s$\\ ($\mu$s)\end{tabular} & \begin{tabular}[c]{@{}c@{}}$T_{aq}$\\ ($\mu$s)\end{tabular} & \begin{tabular}[c]{@{}c@{}}$T_{EH}$\\ ($\mu$s)\end{tabular} & \begin{tabular}[c]{@{}c@{}}$C_{EH}$\\ ($\mu$F)\end{tabular} & \begin{tabular}[c]{@{}c@{}}$V_{EH}$\\ (mV)\end{tabular} & \begin{tabular}[c]{@{}c@{}}$T_{C_{EH}}$\\ (ms)\end{tabular} &  \begin{tabular}[c]{@{}c@{}}$\eta_v$\\ (\%)\end{tabular} & \begin{tabular}[c]{@{}c@{}}$\eta_e$\\ (\%)\end{tabular} \\ \hline
\multirow{2}{*}{\textbf{Sampling@10 KHz}} & \textbf{Schematic} & \multirow{2}{*}{100} & \multirow{2}{*}{10} & \multirow{2}{*}{90} & \multirow{2}{*}{100} & 307.16 & 224.91  & 76.79 & 75.7 \\ \cline{2-2} \cline{7-10} 
 & \textbf{Post-layout} &  &  &  &  & 304.7 & 250.12  & 76.175 & 67 \\ \hline
\multirow{2}{*}{\textbf{Sampling@40 MHz}} & \textbf{Schematic} & \multirow{2}{*}{0.025} & \multirow{2}{*}{0.0025} & \multirow{2}{*}{0.0225} & \multirow{2}{*}{0.025} & 304 & 0.05832  & 76 & 72.64\\ \cline{2-2} \cline{7-10} 
 & \textbf{Post-layout} &  &  &  &  & 300 & 0.063 & 75 & 65.53 \\ \hline
\end{tabular}
\end{table*}

\subsection{Discussion}
For both low and high frequency simulations, it is observed that the input signal energy can be harvested while maintaining the optimal performance of 8 bit SAR ADC (ENOB $\geq$ 7.75bits at the Nyquist rate). The obtained results for high and low frequency simulations are summarized in Table I.

 
The size of the energy harvesting capacitor, i.e., $C_{EH}$ plays an important role and generally based upon the load to be driven. It can be expressed as
\begin{equation}
 C_{EH}=\frac{I_{load}~T_p}{\Delta V}, \label{31}
\end{equation}
where $I_{load}$ is the load current, $T_p$ is the duration between two consecutive peaks of ripples, and $\Delta V$ is the amount of ripple that can be tolerated by the load. From equation \eqref{31}, it is seen that the small value of $\Delta V$ implies large $C_{EH}$. However, a larger value of $C_{EH}$ implies that more time will be taken by the capacitor to charge up to the peak value. Therefore, the capacitor might not be charged to the maximum input signal amplitude, which implies $\eta_v < 1$ \eqref{1}. Hence, depending upon the load conditions, a suitable value of $C_{EH}$ should be selected. 

Generally, electronic systems have lower ripple tolerance capability, and hence $\Delta V \approx 1$mV has been considered in this paper. In order to have a fair comparison between low and high frequency setup, $C_{EH}$ for both setup have been calculated under similar load condition, i.e., $\Delta V \approx 1$mV. The value of $C_{EH}$ obtained for low and high frequency setup is 100 $\mu$F and 25 nF, respectively. It is observed that $\frac{C_{EH}}{T_{C_{EH}}}$ for both setup is around $0.4\times 10^{-3}$ F/s. However, the value of $V_{EH}$ is 307.16 mV and 304 mV for low and high frequency, respectively. Therefore, slightly lower conversion efficiency is obtained for high frequency setup under similar load condition. It may be noted that the high frequency setup can also attain $V_{EH}~=~307.16$mV (same as of low frequency) if $C_{EH}~=~7$ nF. However, such small value of $C_{EH}$ will provide high amount of ripples ($\Delta V~>~7$ mV).



\begin{figure}[ht]
\includegraphics[scale=0.2]{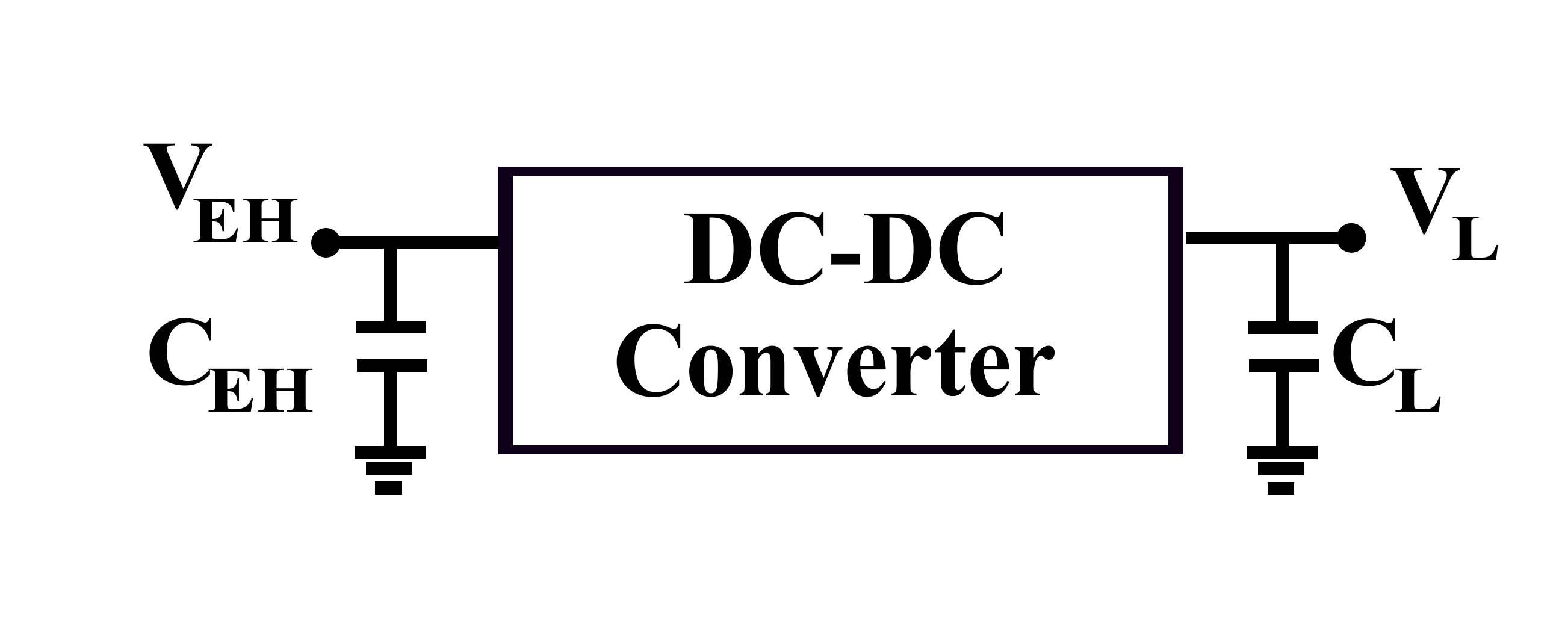}
\centering
\caption{Harvested Voltage boosted to the required level ($V_L$)} 
\label{fig:11}
\end{figure}

Furthermore, if harvested voltage has to drive an application or load which require supply voltage $(V_L)$ much greater than $V_{EH}$, then dc-dc converter \cite{6923477,7387490,6872740} can be deployed to boost $V_{EH}$ up to $V_{L}$ in the EH branch as shown in Fig. \ref{fig:11}.

 \section{Real time Validation of \textit{eSampling}}
 \subsection{Hardware setup}
  \begin{figure}[ht]
\includegraphics[width=\columnwidth, height=8cm]{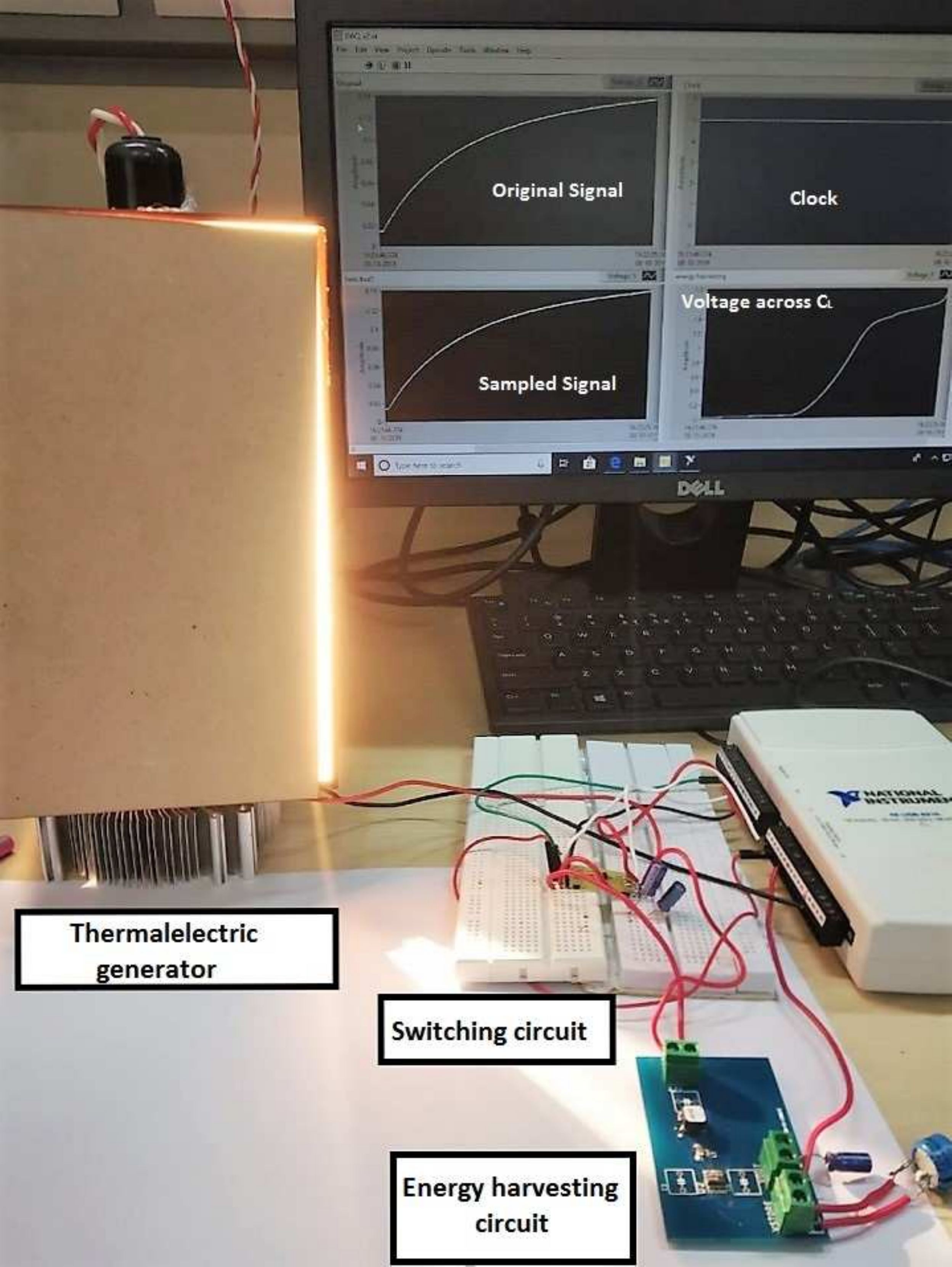}
\centering
\caption{Hardware Prototype} 
\label{fig:16}
\end{figure}
 To demonstrate the proposed \textit{eSampling} framework in real world environment, a hardware set up similar to Fig. \ref{fig:21} a) has been built. The hardware prototype has been shown in Fig. \ref{fig:16}. The input analog signal, $V_{in}(t)$ is obtained via the sensing of the external temperature using a sensing element i.e., thermal electric generator (TEG). TEG can produce an output voltage from 10 mV to 50 mV per $^{\circ}$C change in temperature \cite{haug2017wireless}. The switches $S1$ and $S2$ have been implemented using single pole double throw (SPDT) ADG791 monolithic CMOS switch. A $Clk$ signal similar to the \textit{low-frequency simulation} set up has been generated externally and applied to ADG719. The switch will connect the $V_{in}(t)$ to one output terminal for duration of 10\% of the total sampling period ($T_s$) and to another output terminal with duration of 90\% of the total sampling period, respectively. The output terminals obtaining 10\% and 90\% time of the total sampling period has been connected to $C_h$ for acquiring data samples and $C_{EH}$ for storing energy, respectively. The maximum voltage received at energy storage capacitor i.e., $V_{EH}$ is boosted up to 5V using a dc-dc converter i.e., LTC3108.
\begin{figure}[ht]
\includegraphics[scale=0.57]{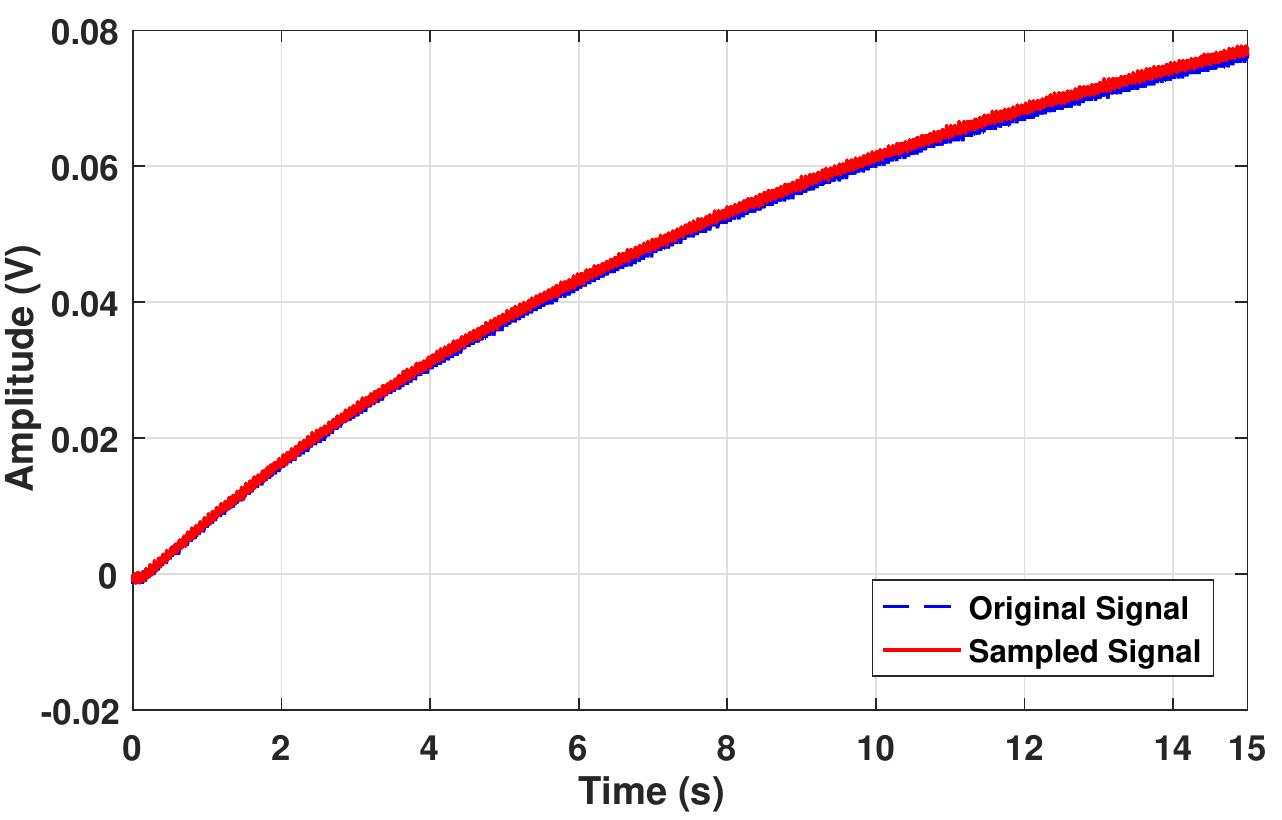}
\caption{Voltage obtained across holding capacitor, $C_{h}$ at 10 KHz sampling frequency in hardware setup .} \label{fig:12}
\end{figure}

\begin{figure}[ht]
\includegraphics[scale=0.57]{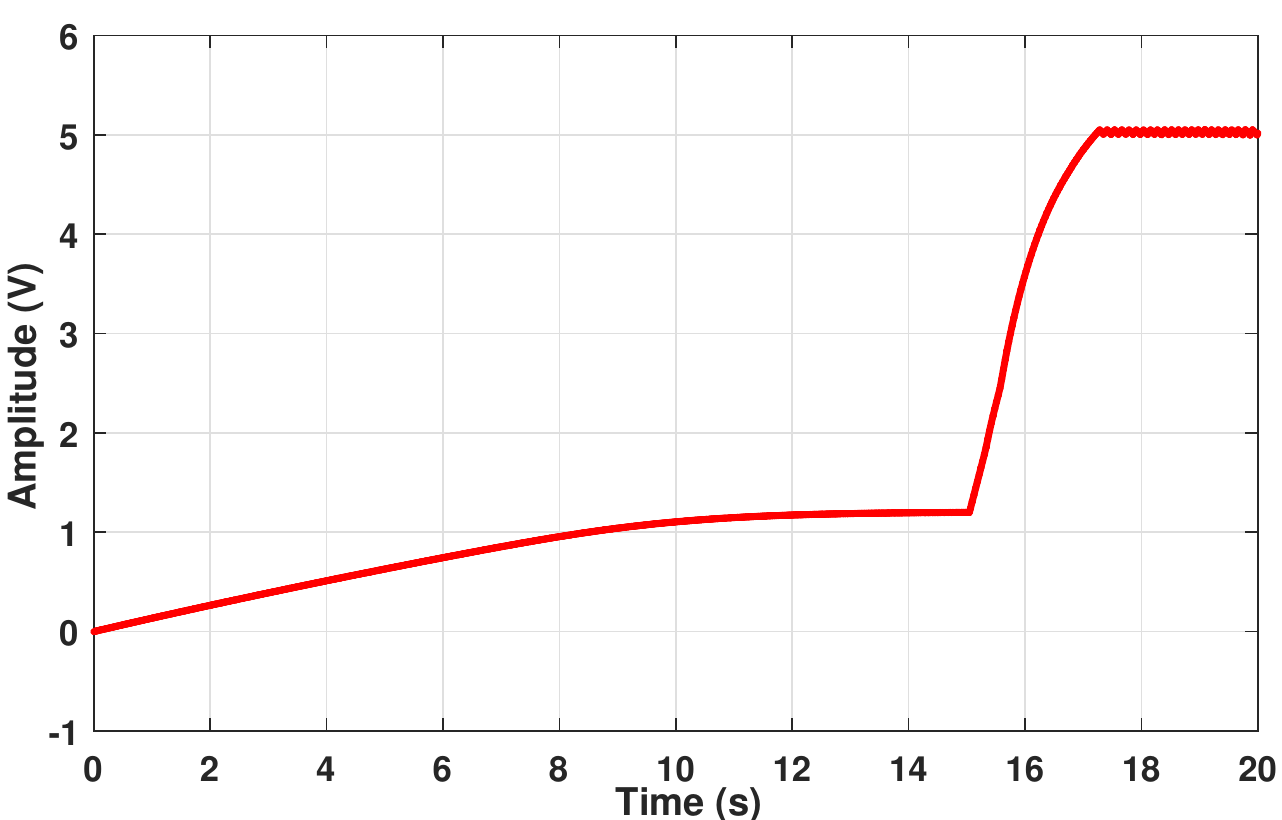}
\caption{Voltage obtained across $C_{L}$ at 10 KHz sampling frequency in hardware setup.} 
\label{fig:13}
\end{figure} 
\subsection{Experimental Results}
Fig. \ref{fig:12} compares $V_{in}(t)$ with the sampled output voltage obtained at $C_h$ using \textit{eSampling}. From the figure, it can be observed that the input signal is similar to the sampled output signal, and hence harvesting the energy along with sampling using \textit{eSampling} is not degrading the performance of conventional sampling process of an ADC. The experiment has been computed for $f_s~=~10$ KHz, $T_{aq}~=~0.1~T_{s}$, and hence $T_{EH}~=~0.9~T_{s}$. Furthermore, the harvested voltage has been boosted to $V_{L}~=~5$V by using LTC3108 as shown in Fig. \ref{fig:13}. The remaining parameters used are as follows: $C_{h}~=~1~\mu$F, $C_{EH}~=~100~\mu$F and $C_{L}~=~10~\mu$F.  The time required by $C_L$ to reach 5V is 16.58 seconds as shown in Fig. \ref{fig:13} with conversion efficiency of 64.98\%.

\section{Conclusion}

In this paper, modifications in  conventional sampling circuitry of an ADC has been proposed for harvesting energy from the discarded portion of the input analog signal. The method is named as \textit{eSampling}. For validating the presented method, CMOS 65 nm technology based system has been designed. In the experiment, only $10\%$ of the sampling period has been used to sample the signal (with perfect recovery of the original analog signal subsequently), while rest $90\%$ has been used for energy harvesting. The experiment has been performed at both low (10 KHz) as well as high frequency (40 MHz). It is observed that more than 65\% conversion efficiency has been obtained with post-layout simulations for both experiments. The experiment has also been replicated on a hardware set up by obtaining the analog signal from TEG at 10 KHz sampling frequency. The obtained conversion efficiency is 64.98\%.
 
 \bibliography{sampling_EH}
 \bibliographystyle{IEEEtran}
\end{document}